\documentstyle[aps,preprint,prl,epsfig]{revtex} 
\newcommand{\up}{\uparrow}
\newcommand{\dn}{\downarrow}
\begin{document}              

\title{Frustrated Hubbard ladders and superconductivity
in  $\kappa$--BEDT--TTF organic compounds }
\author{ S.\ Daul $^{1}$ and D. J. Scalapino $^{2}$  }
\address{ 
$^1$ Institute for Theoretical Physics, University of California, \\
                  Santa Barbara CA 93106. \\
$^2$ Physics Department, University of California, \\
                  Santa Barbara CA 93106. \\
}
\maketitle

\begin{abstract}              

Half--filled two--leg Hubbard ladders have spin--gapped short--range antiferromagnetic
correlations while three--leg ladders have power law antoferromagnetic 
correlations, and both systems have $d_{x^2-y^2}$--power law pairing correlations
when they are doped.
Thus these ladders exhibit some of the phenomenology seen in the layered 
cuprates. Here we report results for half--filled frustrated Hubbard ladders,
based upon ladder segments taken from a tight--binding model of 
$\kappa$--BEDT--TTF.
Although these ladders are half--filled, varying the degree of frustration
can drive them across an insulator--metal transition.
We suggest that the spin, charge and pairing correlations of these frustrated
ladders near the insulator--metal transition provide support for the notion
that $\kappa$--BEDT--TTF is a strongly correlated superconductor.

\end{abstract}

\vspace{10mm}

The organic $\kappa$--BEDT--TTF crystal consists of weakly coupled two--dimensional
layers of BEDT--TTF molecules separated by insulating layers of anions such as
Cu[N(CN)$_2$]Br.
Under pressure, and at low temperature ($T \stackrel{<}{\sim} 10 K$), this 
system undergoes a phase transition from an insulating antiferromagnetic phase
to a superconducting phase.
Based upon parameters obtained from quantum chemistry calculations, it is 
believed that this material is a strongly correlated electron system. 
\cite{McKenzie99,Fortunelli97} 
Within this framework, one can seek to understand the physics of the low 
temperature $\kappa$--BEDT--TTF phase diagram in terms of a model containing
a strong on-site Coulomb interaction and a tight binding parameterization of the
relevant band structure.

A diagram showing the arrangement of the $\kappa$--BEDT--TTF molecules and their
dominant hopping integrals, as proposed by Kino and Fukuyama \cite{Kino96},
is shown in Fig. 1. 
In this model each lattice site, denoted by a circle, contains two BEDT--TTF
molecules which are strongly hybridized leading to a large splitting
between their bonding and antibonding orbitals.
The three electrons on each site leave the antibonding orbital with one 
electron and it is the resulting half--filled $t-\tau$ antibonding band with 
a strong on-site Coulomb interaction $U$ that has been proposed as a minimal
model. For $\tau=0$, this is just a near neighbor $2D$ Hubbard model which
at half--filling has an insulating antiferromagnetic ground state.
When $U$ is large compared to the bandwidth, and $\tau$ increases, the
diagonal exchange leads to frustration of the antiferromagnetic
order and the possibility of a spin gapped phase.
Alternatively, for smaller values of $U$, the system is expected to go from
an insulating antiferromagnetic phase to a paramagnetic metal and possibly
superconducting phase as $\tau$ increases. \cite{McKenzie99}

Motivated by this scenario and the geometry of the tight binding lattice,
we report results obtained from density matrix renormalization group 
\cite{White92} (DMRG) calculations on the two different ladders indicated by
the large arrows shown in Fig. 1. The horizontal ladder, 
corresponding to a one--dimensional chain with a near neighbor hopping $t$
and a next near neighbor hopping $\tau$ has been previously studied 
\cite{Daul99}
and we will summarize its properties in our conclusion.

The diagonal ladder of Fig. 1 has a Hamiltonian
\begin{equation}
 H = -t \sum_{<ij>,\; \sigma} \left( c^\dag_{i,\sigma} c_{j,\sigma} + \mbox{h.c}  \right)
-\tau \sum_{[ij],\; \sigma} \left( c^\dag_{i,\sigma} c_{j,\sigma} + \mbox{h.c}  \right)
+ U \sum_i n_{i\uparrow} n_{i\downarrow}
\label{eq:ham}
\end{equation}
where $c^\dag_{i,\sigma}$ $(c_{i,\sigma}$) creates (annihilates) an
electron on site $i$ with spin $\sigma$. The sum $<ij>$ runs over pairs 
of nearest neighbor sites, and the sum $[ij]$ runs over pairs of 
lattice along only one diagonal of the ladder.
In the following we measure energies in units of $t$,
and the diagonal hopping $\tau$ is positive. 
To investigate this system we applied the density--matrix renormalization
group \cite{White92} technique for system up to size 
$L_y \times L_x = 2\times 32$ with open 
boundary conditions.
We kept up to 800 states so that the maximum discarded weight was
$10^{-6}$. 

The charge gap 
\begin{equation}
   \Delta_c = \frac{1}{2} \left[ E_0(N+2,S=0) + E_0(N-2,S=0) - 
          2 E_0(N,S=0) \right]
\end{equation}
and the spin gap
\begin{equation}
   \Delta_s = E_0(N,S=1) - E_0(N,S=0) 
\end{equation}
for a half--filled ($N=64$) $2 \times 32$ ladder are shown in Figs. 2a and
2b, respectively.
Here $E_0(N,S)$ is the ground--state energy for a state with $N$ electrons
and spin $S$. 
The gaps are plotted versus $U$ for different values of $\tau$.
They were determined by finite--size scaling of the result from systems of
size $L_x$ = 8, 16, 24 and 32. 
The charge gap is set by some fraction of $U$ and hence is relatively
easy to determine. 
The spin gap, which is set by the effective exchange interaction, is small
even for the unfrustrated $\tau=0$ ladder.
When $\tau$ is finite, the resulting frustration further reduces the spin gap.
Hence for small values of $U$, there are strong finite size effects which 
make it difficult to determine the spin gap very precisely. 
But here we are interested in the qualitative phase diagram rather than
quantitative.
We have also calculated the charge and spin gaps for $\tau=1.2$ and $1.5$, 
which were both finite for small $U$. 

For $\tau=0$, we have a half--filled 2--leg Hubbard ladder and as expected
both the charge and spin gaps open up as $U$ turns on.  In opposite, for
$ 0.7 \stackrel{<}{\sim} \tau \stackrel{<}{\sim} 1.1$, 
a finite value of $U$ is required to open up a charge gap.
For example, for $\tau=1$, a value of $U\approx 4$ is required before
the system develops a charge gap and becomes insulating.
When $\tau$ is present, it leads to a frustrating diagonal exchange interaction.
For a sufficiently large value of $\tau$ ($\tau \sim 1$), the dominant singlets are
formed along the $\tau$ diagonal bond rather than the $t$ leg and rung bonds
of the ladder.
Based upon the results of Fig. 2, a schematic phase diagram for the ``diagonal''
ladder is shown in Fig. 3a.

The Hamiltonian of the ``horizontal'' ladder shown in Fig. 1 is identical
to that of the ``diagonal'' ladder for $\tau=1$. 
However, for $\tau \neq 1$, the Hamiltonians for the two ladder differ. 
In particular, for $\tau=0$, the ``horizontal'' ladder reduces to the
1--leg Hubbard model which at half--filling has a charge gap
but no spin gap as opposed to the diagonal ladder which reduces to the
2--leg Hubbard ladder with a spin gap.
Thus for small values of $\tau$, the ``horizontal'' ladder is characterized
by power law antiferromagntic correlations and a charge gap.
The phase diagram for the ``horizontal'' ladder obtain in Ref. \onlinecite{Daul99},
is shown in Fig. 3b. 
In this case one has an insulating phase with power law antiferromagnetic
correlations at smaller values of $\tau$ and a spin gapped insulator with
short range antiferromagnetic correlations separated from a smaller $U$
metallic phase at larger values of $\tau$.

In order to explore the tendency for pairing, we have measured the 
rung--rung singlet pair field correlation function
\begin{equation}
  D(\ell) = \langle \Delta_{i+\ell} \Delta_i^\dag \rangle .
\end{equation}
The operator
\begin{equation}
     \Delta^\dag_{i}  = c^\dag_{i1,\up} c^\dag_{i2,\dn} -
                          c^\dag_{i1,\dn} c^\dag_{i2,\up}
\end{equation}
creates a singlet pair on the $i^{\mbox{{\footnotesize th}}}$ rung and 
$\Delta_{i+\ell}$ 
destroys it on the $ (i + \ell) ^{\mbox{{\footnotesize th}}}$ rung.
In Fig. 4 we show results for  $D(\ell)$ versus $\ell$ for $\tau=1$
and values of $U$ above and below the insulator--metal transition.
For $U=6$, the system is in the insulating phase and $D(\ell)$ decays 
exponentially.
When $U$ decreases below 4, the charge gap vanishes and the pair field
correlations exhibit a power law decay.

In summary, DMRG calculations for 2--leg ladder segments of the model
$\kappa$--BEDT--TTF lattice give phase diagrams which have features similar
to the proposed phase diagram of the $2D$ system. \cite{McKenzie99}
In particular, for small values of $\tau$, they both have an insulating
phase with antiferromagnetic correlations.
In the diagonal ladder, these correlations are short range and the system
is spin gapped while for the horizontal ladder these antiferromagnetic
correlations can be power law or short range depending upon the size of $\tau$.
In addition for moderate to smaller values of $U$, both ladders can go into
a metallic state as $\tau$ increases and this state exhibits power law
singlet pairing correlations.
Thus one could imagine that as the pressure is increased the system crosses
from an insulating state with antiferromagnetic correlations to a metallic
state with singlet pairing correlations.

The authors would like to acknowledge useful discussions with R. H. McKenzie
and S. R. White.
S. Daul acknowledges support from the Swiss National Science Foundation and the
ITP under NSF grant PHY94--07194. 
D. J. Scalapino wishes to acknowledge partial support from the US Department of
Energy under Grant No. DE--FG03--85ER45197.



\begin{figure}
 \begin{center}
 \epsfig{file=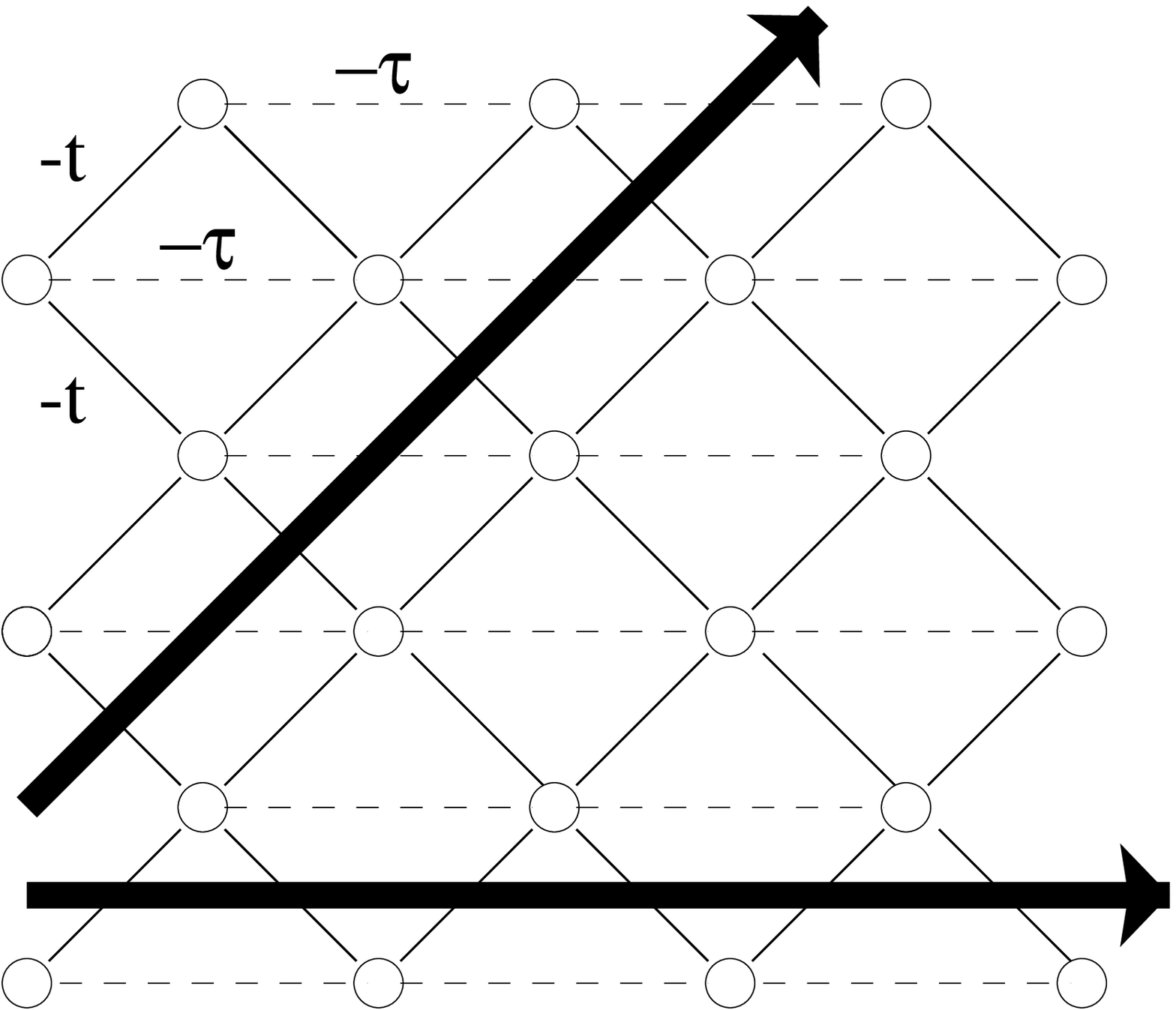,width=6cm}
\caption{Triangular lattice for the dimer model of $\kappa$--(BEDT--TTF)$_2$X.
Each lattice site is denoted by a circle and  represents the anti--bonding 
orbital on a dimer of a pair of BEDT--TTF molecules. 
The large arrows show the two possible ladders.}
 \end{center}
\end{figure}

\begin{figure}[htb]
 \begin{center}
  \epsfig{file=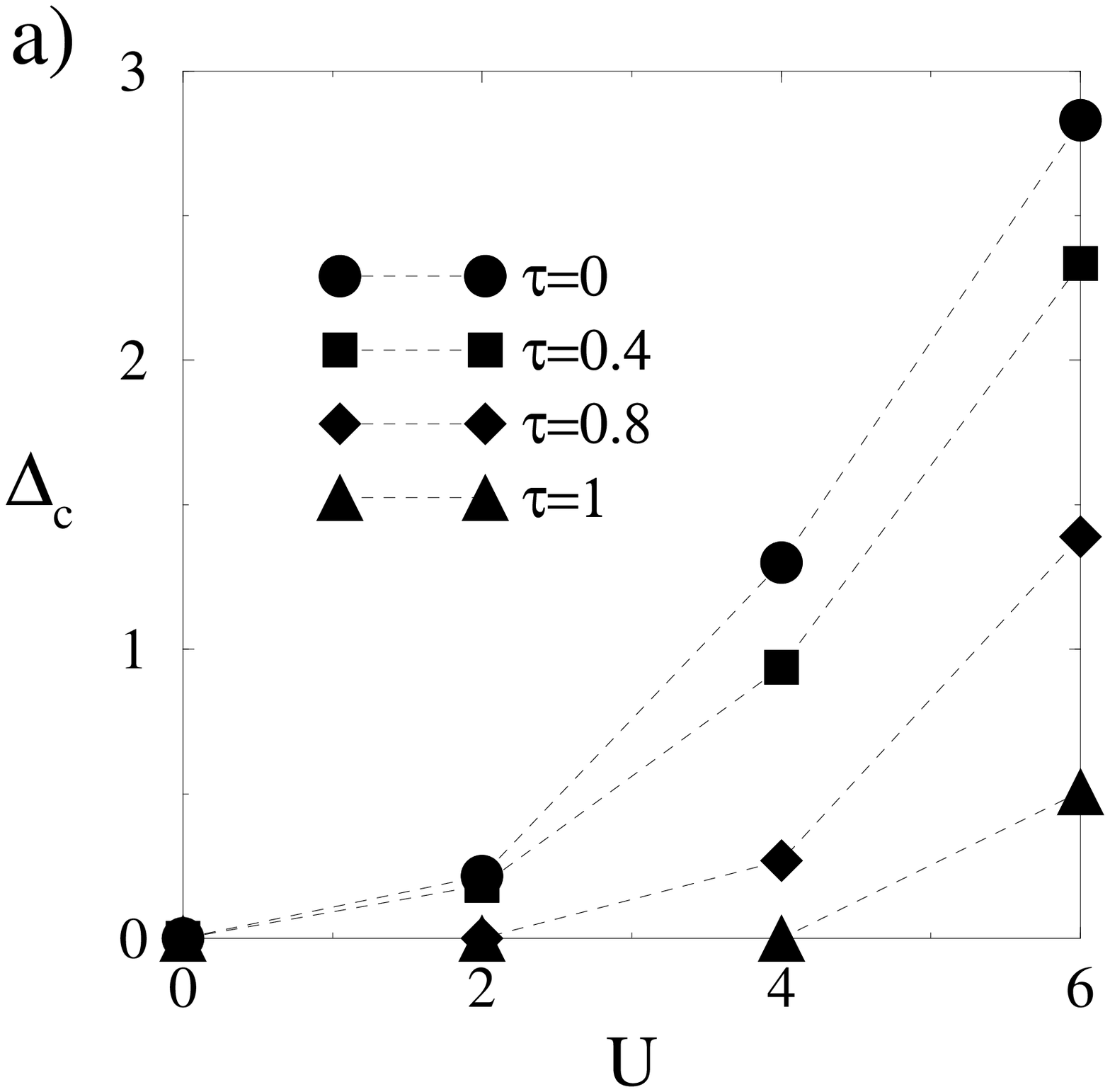,width=60mm}
  \epsfig{file=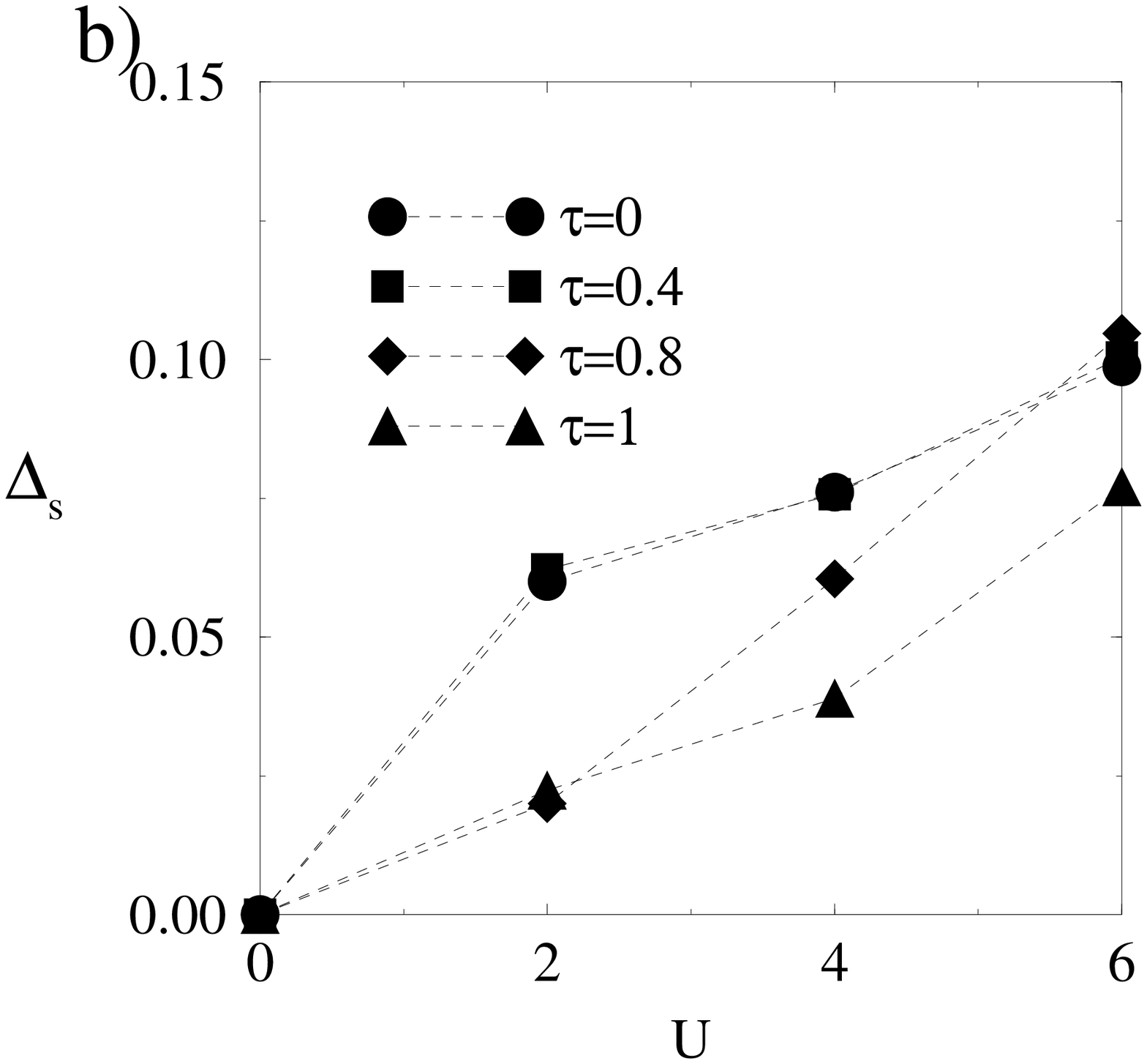,width=65mm}
 \caption{ a) The charge gap of the half--filled ladder as a function of $U$
for various values of $\tau$. 
b) The spin gap of the half--filled ladder as a function of $U$
for  various values of $\tau$.}
 \end{center}
\end{figure}

\begin{figure}
 \begin{center}
 \epsfig{file=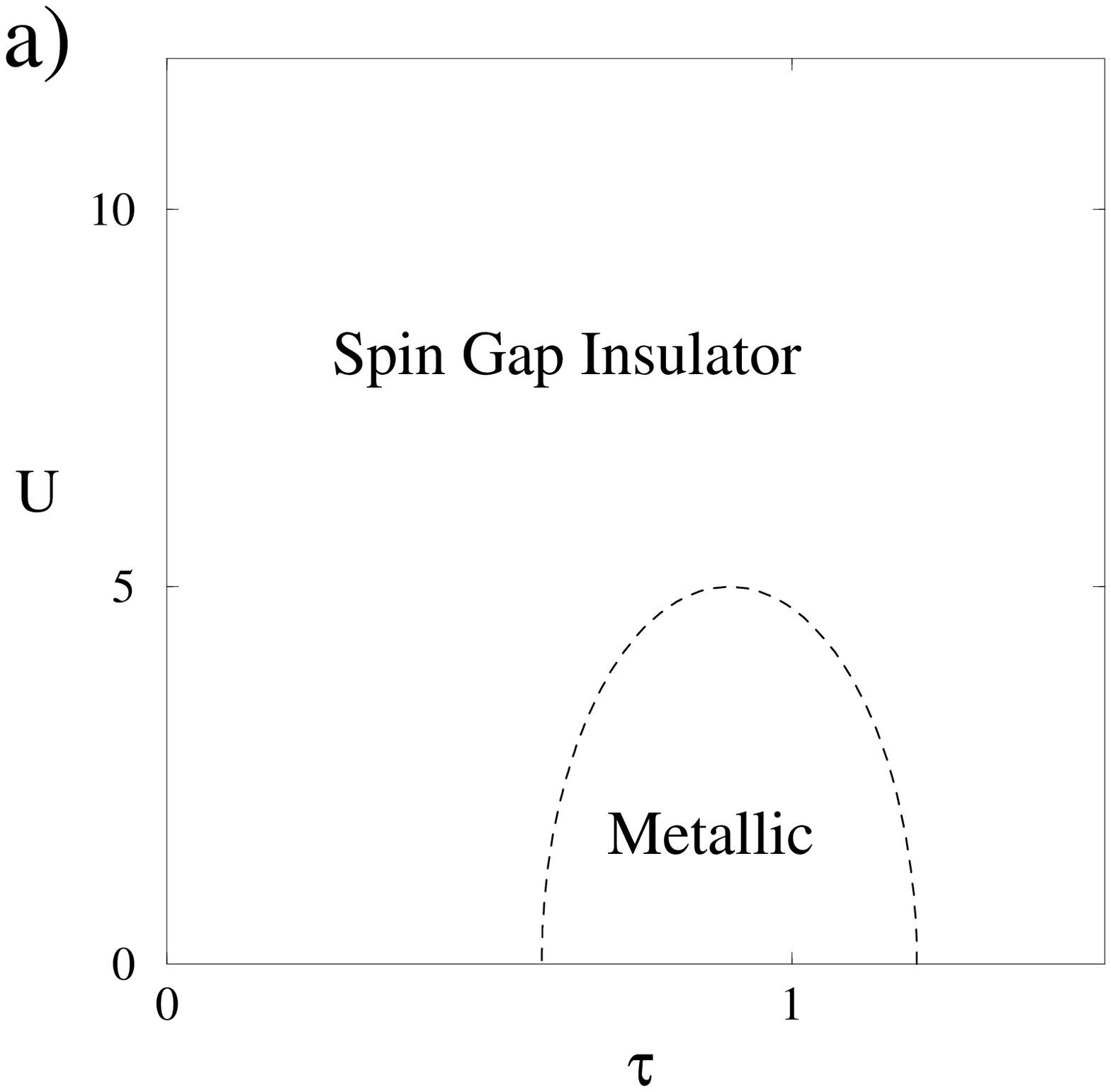,width=6cm}
 \epsfig{file=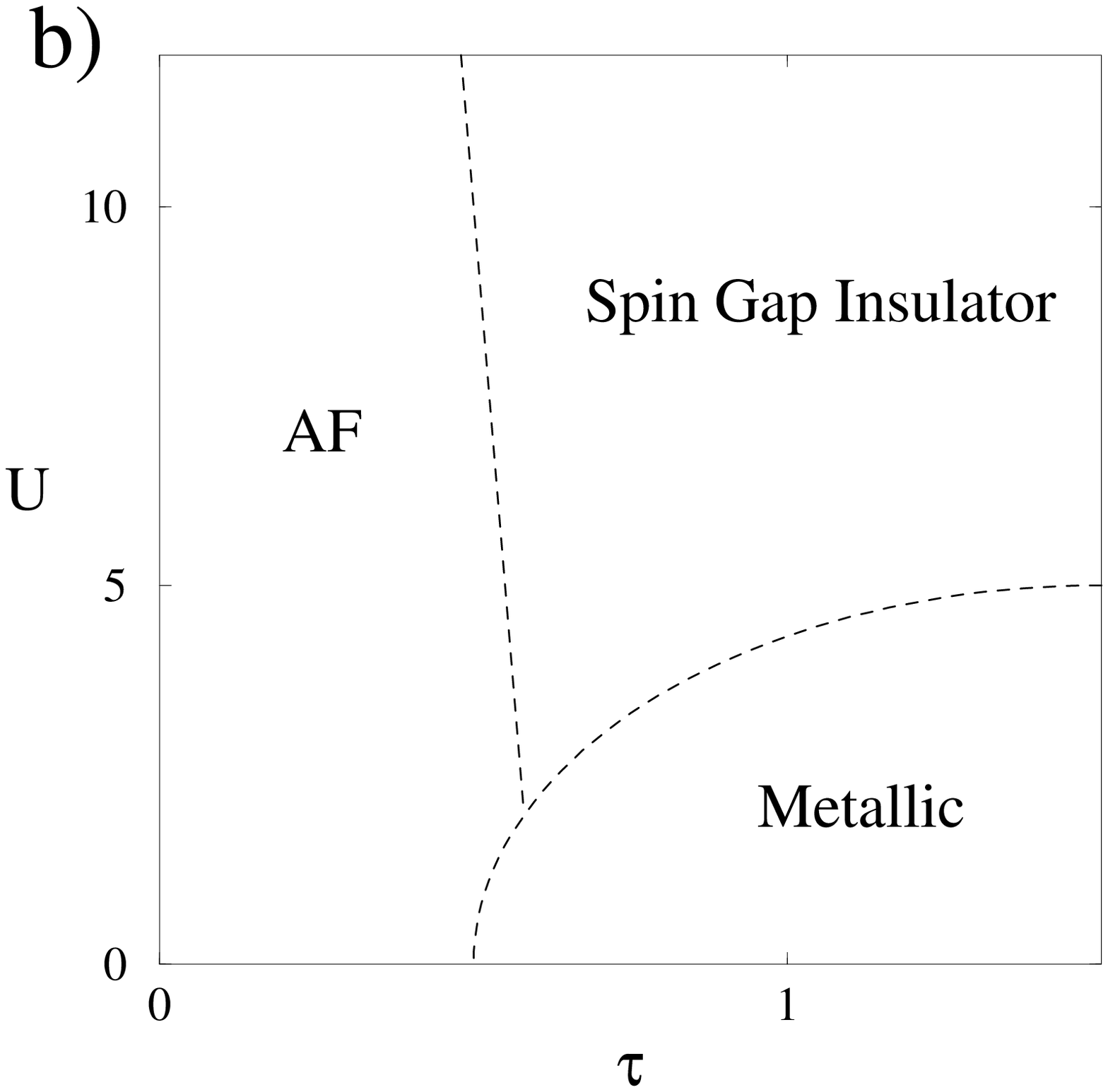,width=6cm}
\caption{Schematic phase diagram of the ``diagonal'' ladder (a) and the 
``horizontal'' ladder (b).}
 \end{center}
\end{figure}

\begin{figure}[htb]
 \begin{center}
  \epsfig{file=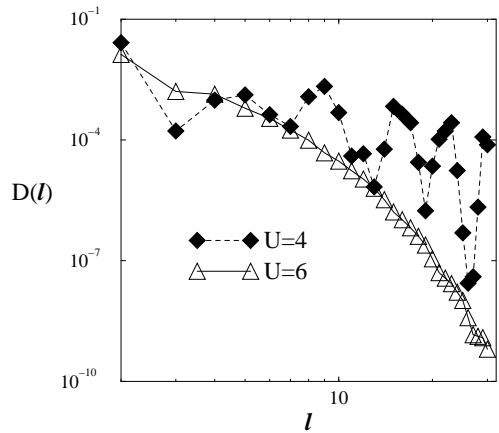,width=64mm}
 \caption{ The rung--rung singlet pairing correlation function
$D(\ell)$ versus $\ell$ for the half--filled ladder
for $\tau=1$ and various values of $U$. }
 \end{center}
\end{figure}

\end{document}